\documentstyle[seceq,preprint,epsfig]{jpsj}
\newcommand{\bc}{\begin{center}}
\newcommand{\ec}{\end{center}}
\newcommand{\be}{\begin{equation}}
\newcommand{\ee}{\end{equation}}
\newcommand{\beqn}{\begin{eqnarray}}
\newcommand{\eeqn}{\end{eqnarray}}
\newcommand{\up}{$|\uparrow\rangle$\ }
\newcommand{\down}{$|\downarrow\rangle$\ }

\def\runtitle{Ground State Structure of Spin Glasses}
\def\runauthor{Matteo {\sc Palassini} and A. P. {\sc Young}}

\title{
Ground State Structure of Spin Glasses
}

\author{Matteo {\sc Palassini} and A. P. {\sc Young}}

\inst{Department of Physics, University of California, Santa Cruz, 
CA 95064}

\recdate
{
\today
}

\abst
{
We investigate the ground state structure of Ising spin glasses in zero
magnetic field by determining how the ground state changes in a fixed finite
block far from the boundaries when the boundary conditions are changed. We
find, both in two and three dimensions, that the probability of a change in the
block ground state configuration tends to zero as the system size tends to
infinity.  This indicates a trivial ground state structure, as predicted by the
droplet theory.
}

\kword
{
Spin glasses, ground states, optimization methods.
}

\begin{document}
\sloppy
\maketitle

\section{Introduction}
In this talk we discuss some recent work that tries to shed light on
the nature of the spin glass state. It is now fairly clear that there is a
finite temperature spin glass transition in three
dimensions\cite{ky,mari3d,hart,matteo,other-theta-3d}
(3-$d$) though a transition only occurs\cite{rieger,other-theta} at
$T=0$ in 2-$d$. However, the nature of the
spin glass phase below the transition temperature $T_c$ is not clear. We will
review the two principal scenarios that have been proposed for the spin glass
phase, which differ in how many ``states'' contribute to the correlation
functions, and will mention briefly existing
numerical results which probe this region. Then we describe a
new approach\cite{py,py2}
to investigate the problem which involves looking at how sensitive 
the region far from the boundaries is to changes in the
boundary conditions. The idea can be applied both to ground states and to
finite temperature studies, but initially we have just considered $T=0$.

\section{Scenarios for the spin glass phase}
Controversy remains over the nature of ordering in spin glasses below the
transition temperature, $T_c$, and two scenarios have been extensively
discussed.

In the first approach, one assumes that the basic structure of the replica
symmetry breaking (RSB) in Parisi's\cite{parisi,mpv,by,mari99} solution of the
infinite range model applies also to realistic short range systems. In this
picture, the order parameter is not just a single number, as is the case for a
ferromagnet for example, but is a probability distribution, $P(q)$. This
distribution gives
the probability that two spin configurations, weighted by the Boltzmann
factor, have overlap given by $q$.
To be precise $P(q)$ is defined by
\begin{equation}
P(q) = \langle \delta \left( q - q_{12} \right) ,
\end{equation}
where $\langle \cdots \rangle$ denotes both a thermal average and an average
over disorder, and
\begin{equation}
q_{12} = {1 \over N} \sum_{i=1}^N S_i^{(1)} S_i^{(2)} , 
\end{equation}
where $(1)$ and $(2)$ denote two copies of the system with the same
interactions.

In the Parisi
picture $P(q)$ is a non-trivial function because many thermodynamic states
contribute to the partition function, i.e. they have differences in total
(free) energy which are of order unity and yet have very different spin
configurations from each other.  As a result, $P(q)$ has a delta function at
$q_{EA}$ coming from ordering in a single state, and a tail down to $q=0$ from
overlap between different states, which does not vanish for $L \to \infty$.

In the alternative approach, the ``droplet model'', proposed by Fisher and
Huse\cite{fh} (see also Refs.~\cite{bm,mcmillan,ns-old,ns-new}), 
one starts by defining the concept of ``thermodynamic states'' and
``pure states''. For a given set of boundary conditions, one looks at the
correlation functions of the spins in the bulk, i.e. in a 
finite region far from the
boundary.  Each different set of correlation functions defines a separate
thermodynamic state.

However, many thermodynamic states defined in this way are
related to each other. To see this, consider,
for example, a ferromagnet below $T_c$. Clearly two
of the
thermodynamic states are \up and \down, where the
spins are aligned up and down respectively. These can be generated by fixing
the spins on the boundary to be up (or down). However, if we use
{\em periodic}
boundary conditions, then the system is in a linear combination of
\up and \down with equal weight (assuming no
external field, which will be case for all the discussions in this paper).
States like \up and \down, which are ``extremal'', in the sense that other
states can be expressed as linear combinations of them, are called pure states.
An important question is the number of
these pure states. For the ferromagnet, there
are just two.
In such a situation, where there is just one pure state (plus
other(s) related by a global symmetry of the Hamiltonian),
we say that the pure state structure is
trivial. If this occurs at $T=0$ we refer to a ``trivial ground state
structure''.

One other issue relating to the ferromagnet needs to be addressed before we go
on to spin glasses. It is possible that the boundary conditions
generate a ``domain'' state which is \up in some region of space and \down in
the rest. If the domain wall intersects the region where we are computing the
correlation functions, we would have another pure state, since it is
not a linear combination of \up and \down (more precisely it is one linear
combination in part of the region and a different linear combination in the
rest). Since these  domain states are closely
related to the pure states \up and \down we will
still denote the pure state structure as trivial. In order to eliminate
domain states, which do not really alter the nature of the ground state
structure, but still enable us to detect
other possible states,
we look at correlation functions in a fixed
{\em finite} box, far from the boundaries, as the (linear)
system size $L$ tends to
infinity. The probability that the domain wall intersects the block vanishes as
$L \to \infty$, and so the ground state structure is trivial if we get
the same state in the box (or the reversed state) with probability one for
$L \to \infty$ when the boundary conditions are changed.

Returning now to spin glasses,
Fisher and Huse\cite{fh} find, under certain assumptions,
that the pure state structure
in spin glasses is trivial. More precisely, they argue that the elementary
excitations are compact objects called ``droplets'' such that the minimum
energy of a droplet of linear extent $L$
which encloses a given site is $\sim L^\theta$, where
$\theta$ is a ``stiffness'' exponent.
For the spin glass state to be stable at finite
temperature one must have $\theta > 0$. Since it costs a lot of energy to turn
over a large number of spins, the state far from the boundary will
not change upon changing the boundary conditions,
as in a ferromagnet where in
that case $\theta = d-1$.  As an additional
consequence, 
$P(q)$ is also trivial, i.e. is a pair of delta functions at $q=\pm q_{EA}$
where $q_{EA}$ is the Edwards-Anderson order parameter describing order in
the (single) pure state. For a finite system,
there will be a weight in the tail but this vanishes in the thermodynamic limit
as $1/L^\theta$.

It should be pointed out that although the volume of a droplet is assumed to be
compact, i.e.  $V \sim L^d$, the surface of the droplets will be a fractal of
fractal dimension, $d_f$ where $d-1 < d_f < d$. Furthermore,
in the droplet picture, the spin glass
state {\em is} much richer than the ferromagnet, even though the pure state
structure is trivial, because the relative spin
orientations at large distances change when the temperature is changed by even
a small amount. This is sometimes called ``chaos''.

Fisher and Huse have also argued that $P(q)$ is not necessarily a good
indicator of the number of pure states. For example an antiferromagnet with an
odd lattice size $L$ must have a domain wall built in. This domain wall will
fluctuate in position and so give a non-trivial $P(q)$, although the pure state
structure is trivial. An opposite example, where $P(q)$ is trivial but there
is more than one pure state is the random field Ising model. The two pure
states are \up and \down, but these
are no longer related by symmetry
and have a (free) energy difference of order $L^{d/2}$. Because this difference
diverges for $L \to\infty$ one state will dominate the statistical sum so
$P(q)$ will just be a single delta function.

\section{Earlier Numerical Work}

First of all we discuss earlier numerical work at zero temperature. In two
dimensions, it is clear\cite{rieger,other-theta}
that $\theta$ is about $-0.28$, the negative value
indicating that spin glass order does not persist to finite temperature because
large domains, which cost very little energy, will be excited at arbitrarily
low temperatures. In three dimensions, $\theta$ is
about\cite{hart,other-theta-3d} 0.20,
which is
positive, implying a finite $T_c$ which has also been found directly.
However, it is also
small which means that it is hard to distinguish the droplet theory
from the RSB
picture from simulations of $P(q)$.
In four dimensions, recent work\cite{theta-4d} has found a fairly large positive
value of $\theta$ of around 0.7. Hence 4-$d$ should be the easiest case in
which to
distinguish droplet from RSB predictions for $P(q)$. 

At finite temperature,
Monte Carlo simulations on short range models on small
lattices in three and four dimensions\cite{rby,marinari,zuliani,berg}, find a
non-trivial $P(q)$ with a weight at $q=0$ which is independent of system size
(for the range of sizes studied), as predicted by the Parisi theory. Some of
the 4-$d$ results show a $P(q)$ which is size independent up to around $L=8$ at
about 2/3 $T_c$, which is surprising if the droplet theory is
correct. Perhaps the droplet theory is correct but there is some length scale,
greater than a lattice spacing, below which the RSB picture works better.
This length could, perhaps, be the
critical correlation length, but naively, it should be quite
small at 2/3 $T_c$ since the correlation length exponent $\nu$ is
less than one \cite{nu-4d}.

\section{New Approach}
Most numerical work up to now
has concentrated on $P(q)$
but here\cite{py,py2}, by contrast, we focus directly on the
the pure state structure.
It is interesting to investigate this even at $T=0$, where
there are 
efficient algorithms for determining ground states, even though
$P(q)$ is 
trivial in this limit (for a continuous bond
distribution).

We look at how the spin correlations 
in a central block of fixed
size $N_B = L_B^d$, change when the boundary conditions
change. Remember, if the probability of a change tends to zero for
$L \to \infty$, then the ground
state structure is trivial. 
Here we just consider $T=0$ and consider a central block of size $L_B=2$.
Note that there is no need for the block size to tend to infinity.
This would be very inconvenient to implement since we need $L_B \ll L$ and yet
the largest size $L$ that can be studied is not extremely large..

According to the droplet theory, the change in boundary conditions will induce
a domain wall of size $L$ plus possibly some smaller domains near the boundary.
The configuration of the central block will change if the the surface of the
droplet of size $L$ passes through the block. The probability that this happens
is proportional to
\begin{equation}
{1 \over L^{d - d_f} } ,
\label{dmdf}
\end{equation}
where $d_f$ is the fractal dimension of the surface of the droplet (i.e. the
domain wall). Hence this probability tends to zero for $L \to \infty$, at least
as long as $d_f < d$.

According to the RSB picture, changing the boundary conditions will shuffle the
order of the low energy states (in general we expect by an amount proportional
to $L^{(d-1)/2}$). Anti-periodic boundary conditions, see below, will have a
smaller effect. Since there are assumed to be states which differ in energy by
a finite amount, and which have very different spin configurations (since the
overlap is less than unity), the new ground state will be quite different from
the old one. Hence the probability that the spin configuration in the block
changes is non-zero for $L \to \infty$.

\section{The Model}

The Hamiltonian is given by
\begin{equation}
{\cal H} = -\sum_{\langle i,j \rangle} J_{ij} S_i S_j ,
\label{ham}
\end{equation}
where the sites $i$ lie on a 
simple cubic ($d=3$) or square lattice ($d=2$) with $N=L^d$ sites 
($L \le 10$ in $3d$, $L \le 30$ in $2d$) , $S_i=\pm
1$, and the $J_{ij}$ are nearest-neighbor interactions chosen according to a
Gaussian distribution with zero mean and standard deviation unity.
We determine the energy and spin configuration of the ground state for a given
set of bonds, initially
for periodic boundary conditions denoted by ``P''.
Next we impose anti-periodic conditions (``AP'') along {\em
one}\/ direction, which is equivalent to keeping periodic boundary conditions
and changing the sign of the
interactions along this boundary, and recompute the ground state. Then we
change the sign of  half the bonds at random along this boundary, which we
denote by ``R''. Finally we replace the bonds on {\em all} the surfaces by a
new set of random variables (so the magnitude as well as the sign is changed).
We denote this by ``R3''. 

In 2-$d$  we used sizes up to $L=30$ while for 3-$d$ the largest size was
$L=10$. 
To determine the ground state in two dimensions we used the Cologne spin glass
ground state server\cite{juenger} which computes {\em exact} ground states
using a branch-and-cut algorithm. In three dimensions we used a heuristic
``genetic'' algorithm discussed by Pal\cite{pal1,pal2}, see also
Refs.~\cite{py2,Gen_Alg_book}.
We did some checks\cite{py2} to verify that errors due to the
genetic algorithm sometimes not giving the exact ground state are negligible.

In order to study the dependence of the spin configurations
in the central block on boundary conditions
we compute
the block spin overlap distribution
$P^B_{\alpha\beta}(q)$, where $\alpha$ and $\beta$ denote two boundary
conditions, and
\begin{equation}
P^B_{\alpha\beta}(q) = \left\langle \delta \left( q - q^B_{\alpha\beta}
\right) \right\rangle ,
\end{equation}
in which
\begin{equation}
q^B_{\alpha\beta} = {1\over N_B} \sum_{i=1}^{N_B} S_i^\alpha S_i^\beta 
\end{equation}
is the overlap between the block configurations with
$\alpha$ and $\beta$ boundary conditions, $S_i^\alpha$ is the value of $S_i$
in the ground state with the $\alpha$ boundary condition, 
and the brackets $\langle \cdots \rangle$ refer to an average over the
disorder. 

Since we work at $T=0$, each sample and pair $\alpha,\beta$
gives a single value for $q$. 
The self overlap distribution, $P^B_{\alpha\alpha}(q)$,
has weight only at $q = \pm 1$,
since the ground state is unique for a given boundary condition.
$P^B(q)$ is normalized
to unity i.e. $\int P^B(q)\, dq = 1$, it is symmetric,
and the allowed $q$-values are discrete with a separation of of
$\Delta q = 2 / N^B$, so
$P^B_{\alpha\alpha}(\pm 1) = {N^B / 4 }. $

If the configuration
in the block changes when the boundary conditions are changed
from $\alpha$ to $\beta$, 
then the block overlap, $q^B_{\alpha\beta}$,
will no longer be $\pm 1$. Hence 
$ 1 -  P^B_{\alpha\beta}(1) / P^B_{\alpha\alpha}(1) $
is the probability that the block ground state changes on changing the boundary
conditions.
We will see that this quantity
vanishes as $L\to \infty$,

\section{Results}


\begin{figure}
\begin{center}
\epsfig{figure=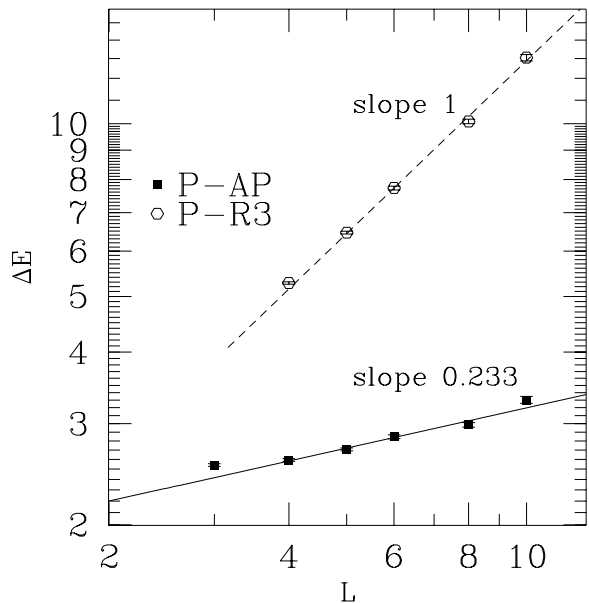,width=8.5cm}
\end{center}
\caption{
Defect energies in 3-$d$.
}
\label{de_rms_3d}
\end{figure}

Fig.~\ref{de_rms_3d} shows results
for the defect energies in 3-$d$ for P-AP and P-R3 boundary conditions.
For P-AP, the boundary condition change can be removed locally by a gauge
transformation\cite{py,py2}
and one is left with a single domain wall, in general far from
the boundary, of size $L$.
For P-AP boundary
conditions we estimate that the data point for $L=10$ is about 5\% too high
because the genetic algorithm does not always find the exact ground state.
Correcting for this, we get $\theta = 0.21 \pm 0.02$ in agreement with other
estimates\cite{hart,other-theta-3d}.
For the P-R3 boundary conditions, the change
cannot be removed locally by a gauge transformation and small
droplets, in the vicinity of the boundary, will also be induced, giving an
energy change of order $L^{(d-1)/2} \sim L$ which agrees with the
numerics.

\begin{figure}
\begin{center}
\epsfig{figure=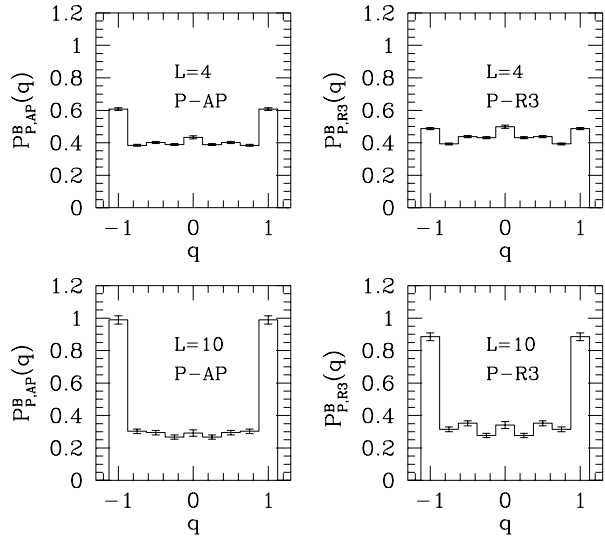,width=8.5cm}
\end{center}
\caption{
Block overlaps in 3-$d$. The probability that the block configuration is
unchanged is the weight in the bins at $q= \pm 1$.
}
\label{hist_multi_3d}
\end{figure}

\begin{figure}
\begin{center}
\epsfig{figure=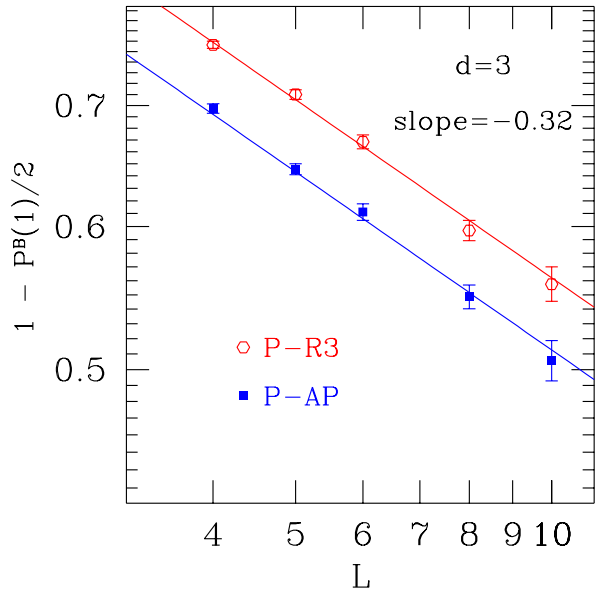,width=8.5cm}
\end{center}
\caption{
Probability that the block configuration changes in 3-$d$ for a range of sizes
on a log-log plot.
}
\label{pq1_3d}
\end{figure}

Fig.~\ref{hist_multi_3d} shows block spin overlaps in 3-$d$ for P-AP and P-R3
boundary conditions for two different sizes. One clearly sees that the peaks at
$q = \pm 1$ (whose weight gives the probability that the configuration did
{\em not} change) increase significantly for larger $L$.
The probability that the block configuration changes when the boundary
conditions are changed is shown in
Fig.~\ref{pq1_3d} for 3-$d$ for P-AP and P-R3 changes.
The data for P-R boundary conditions is similar but shows some corrections for
smaller sizes\cite{py2}. The data in Fig.~\ref{pq1_3d} fit straight lines with
the same slope for the whole range of sizes. Fitting to the form
$a  + b L^{-\lambda}$, the chi-squared is higher for any positive $a$ than for
$a = 0$. However, we cannot definitely rule out that the data might extrapolate
to a small positive value for $L \to \infty$.

In 3-$d$ data for both sets of boundary conditions is consistent with the
probability of a change in the block configuration tending to zero as
$L^{-\lambda}$ with $\lambda \equiv d - d_f = 0.32\pm 0.02$. These results
imply that the ground state structure is trivial in Ising spin glasses in three
(and also in two) dimensions. We believe that our value for $d_f$ is the first
reliable prediction for this quantity in 3-$d$.  
The data for 2-$d$ shows\cite{py,py2} good power law behavior
according to Eq.~(\ref{dmdf})
with $d - d_f = 0.69 \pm 0.02$, in good agreement
with Middleton\cite{midd} and earlier estimates of
$d_f$\cite{rieger,other-df,gingras}.

\section{Conclusions}
To conclude we have seen that the ground state structure appears to be trivial
in a spin glass model with a finite $T_c$, the three-dimensional Ising spin
glass.  It remains to understand why
Monte Carlo simulations at finite 
temperature find,
by contrast,
evidence for a non-trivial pure state structure. It seems likely either that
the droplet theory is correct and a trivial $P(q)$ is only seen for larger
sizes at finite-$T$, or, possibly, that
there are other low energy excitations which are not
seen by changing the boundary conditions and which could give a non-trivial
$P(q)$ in the thermodynamic limit.

\section*{Acknowledgments}
We should like to thank Daniel Fisher, Marc M\'ezard, Jean-Philippe Bouchaud
and Giorgio Parisi for helpful comments.
This work was supported by the National Science Foundation under grant DMR
9713977.  M.P. is supported in part by University of California, EAP Program,
and by a fellowship of Fondazione Angelo Della Riccia. The numerical
calculations were made possible by allocations of time from 
the National Partnership for 
Advanced Computational Infrastructures (NPACI) and
the INFM Parallel Computing Initiative.
We also thank Prof. M.~J\"unger and his group
for putting their spin glass ground state server in the
public domain.


\begin{thebibliography}{99}
\bibitem{ky}
    N.~Kawashima and A.~P.~ Young, Phys. Rev. B {\bf 53}, R484, (1996).

\bibitem{mari3d}
    E.~Marinari, G.~Parisi, and J.~J.~Ruiz-Lorenzo Phys. Rev. B. {\bf 58},
    14852 (1998)

\bibitem{matteo}
    M.~Palassini and S.~Caracciolo, Phys. Rev. Lett, {\bf 82}, 5128 (1999).

\bibitem{hart}
   A.~K.~Hartmann, Phys. Rev. E {\bf 59}, 84 (1999).

\bibitem{other-theta-3d}
    A.~J.~Bray and M.~A.~Moore, J. Phys. C, {\bf 17}, L463 (1984);
    W.~L.~McMillan, Phys. Rev. B {\bf 30}, 476 (1984); M.~Cieplak and
    J.~Banavar, J. Phys. A {\bf 23}, 4385 (1990).

\bibitem{rieger}
     H.~Rieger, L.~Santen, U.~Blasum, M.~Diehl, and M.~J\"unger, J. Phys. A {\bf
     29}, 3939 (1996).

\bibitem{other-theta}
    A.~J.~Bray and M.~A.~Moore, J. Phys. C {\bf 17}, L463 (1984);
    W.~L.~McMillan, Phys. Re. Rev. B {\bf 29}, 4026 (1984); M.~Cieplak and
    J.~R.~Banavar, J. Phys. A {\bf 23}, 4385 (1990); D.~A.~Huse and L.-F. Ko,
    Phys. Rev. B {\bf 56}, 14597 (1997).

\bibitem{py}
    M.~Palassini and A.~P.~Young, Phys. Rev. B {\bf 60}, R9919 (1999).

\bibitem{py2}
    M.~Palassini and A.~P.~Young, cond-mat/9906323

\bibitem{fh}
    D.~S.~Fisher and D.~A.~Huse, J. Phys. A. {\bf 20} L997 (1987); D.~A.~Huse
    and D.~S.~Fisher, J. Phys. A. {\bf 20} L1005 (1987); D.~S.~Fisher and
    D.~A.~Huse, Phys. Rev. B {\bf 38} 386 (1988).

\bibitem{bm}
    A.~J.~Bray and M.~A.~Moore, in {\em Heidelberg Colloquium on Glassy
    Dynamics and Optimization}\/, L.~Van~Hemmen and I.~Morgenstern eds.
    (Springer-Verlag, Heidelberg, 1986).

\bibitem{mcmillan}
    W.~L.~McMillan, J. Phys. C {\bf 17} 3179 (1984).

\bibitem{ns-old}
    C.~M.~Newman and D.~L.~Stein, Phys. Rev. B {\bf 46}, 973 (1992); Phys. Rev.
    Lett., {\bf 76} 515 (1996).

\bibitem{ns-new}
    C.~M.~Newman and D.~L.~Stein, Phys. Rev. E {\bf 57} 1356 (1998).


\bibitem{parisi}
    G.~Parisi, Phys. Rev. Lett. {\bf 43}, 1754 (1979); J. Phys. A {\bf 13},
    1101, 1887, L115 (1980; Phys. Rev. Lett. {\bf 50}, 1946 (1983).

\bibitem{mpv}
    M.~M\'ezard, G.~Parisi and M.~A.~Virasoro, {\em Spin Glass Theory and
    Beyond} (World Scientific, Singapore, 1987).

\bibitem{by}
    K.~Binder and A.~P.~Young, Rev. Mod. Phys. {\bf 58} 801 (1986).

\bibitem{mari99}
    E.~Marinari, G.~Parisi, F.~Ricci-Tersenghi, J.~Ruiz-Lorenz and
    F.~Zuliani, cond-mat/9906076.

\bibitem{rby}
    J.~D.~Reger, R.~N.~Bhatt and A.~P.~Young, Phys. Rev. Lett. {\bf 64}, 1859
    (1990).

\bibitem{marinari}
    E.~Marinari, G.~Parisi, and J.~J.~Ruiz-Lorenzo, in {\em Spin Glasses and
    Random Fields}, edited by A.~P.~Young (World Scientific, Singapore, 1998).


\bibitem{theta-4d}
    A.~K.~Hartmann, cond-mat/9904296; K. Hukushima, cond-mat/9903391.
    
\bibitem{zuliani}
    E.~Marinari and F. Zuliani, cond-mat/9904303.   

\bibitem{berg}
    B.~A.~Berg and W.~Janke, Phys. Rev. Lett. {\bf 80}, 4771 (1998).

\bibitem{nu-4d}
    D.~Badoni et al. Europhys. Lett. {\bf 21}, 495 (1993); G.~Parisi,
    F.~R.~Tersenghi and J.~J.~Ruiz-Lorenzo, J. Phys. A {\bf 29}, 7943 (1996).
    K. Hukushima, cond-mat/9903391

\bibitem{midd}
    A.~A.~Middleton, Phys. Rev. Lett. {\bf 83}, 1672 (1999).

\bibitem{other-df}
    A.~J.~Bray and M.~A.~Moore, Phys. Rev. Lett. {\bf 58}, 57 (1987).

\bibitem{gingras}
    Michel~J.~P.~Gingras, Phys. Rev. Lett. {\bf 71}, 1637 (1993).

\bibitem{pal1}
    K.~F.~Pal, Physics A {\bf 223},  283 (1996).

\bibitem{pal2}
    K.~F.~Pal, Physics A {\bf 233},  60 (1996).

\bibitem{Gen_Alg_book}
    Z.~Michalewicz, {\em Genetic Algorithms + Data Structures =
    Evolution Programs}, (Springer-Verlag, Berlin, 1994.)

\bibitem{juenger}
    Information about the Cologne spin glass ground state server
    can be obtained at
    http://www.informatik.uni-koeln.de/ls\_juenger/projects/sgs.html.

\end{thebibliography}
\end{document}